\documentclass[runningheads,dina4,12pt]{stamm}
\usepackage{graphicx}  

\begin{document}
%
\title*{Universal Unfolding of Pitchfork Bifurcations
and Shear-Band Formation in Rapid Granular Couette Flow
\footnote{
Proceedings of the Symposium on Trends in Applications of Mathematics to Mechanics,
22-28 August 2004, Seeheim, Germany (Editors: K. Hutter and Y. Wang;
Shaker Verlag)}
}
\toctitle{Title of a Manuscript Submitted for Publication\protect\newline
in the Proceedings of STAMM 2004}
%
%
\titlerunning{Universal Unfolding and Shearband Formation in Granular Couette Flow}
%
\author{Meheboob Alam}
\authorrunning{Meheboob Alam}
\institute{Engineering Mechanics Unit, Jawaharlal Nehru Center for
Advanced Scientific Research, Jakkur P.O., Bangalore 560064, India;
Email: meheboob@jncasr.ac.in
}

\maketitle              

\begin{abstract}
A  numerical bifurcation analysis
is carried out to understand
the role of gravity on the shear-band formation in
rapid granular Couette flow.
At {\it low} shear rates,
there is a unique solution with a {\it plug} near the bottom wall
and a {\it shear-layer} near the top-wall;
this solution mirrors typical shearbanding-type profiles
in earth-bound shear-cell experiments.
Interestingly, a {\it stable} plug near the top-wall
is also a solution of these equations at {\it high} shear rates;
there is a multitude of other plugged states, with
the plugs being located in an ordered fashion within the Couette gap.
The origin of such shearbanding solutions is tied to the
spontaneous symmetry-breaking {\it shearbanding} instabilities of the
gravity-free uniform shear flow, leading to both subcritical
and supercritical pitchfork bifurcations.
In the language of singularity theory,
we have established that this bifurcation problem admits {\it universal unfolding}
of pitchfork bifurcations.
\end{abstract}

\section{Introduction}

Granular materials (e.g. sand, coal, cereals, powders, etc.) are of
immense importance in many industrial and geological processes;
most agricultural and pharmaceutical products are in granular form.
The motion of a collection of macroscopic solid particles 
represents a granular flow \cite{Hutter1994,HHL1998}
as in a vibrofluidized bed.
The external driving is essential to sustain such flows
since the particle collisions are inherently dissipative, leading to 
a `continual' energy loss.
Thus, granular flows  represent a {\it driven-nonequilibrium} system.
In typical `dry' granular flows, the effect of interstitial fluid 
is neglected and the
interactions between grains are {\it dissipative}, which, in turn,
leads to a wealth of interesting behaviour \cite{Hutter1994,HHL1998}.
Despite their practical importance and non-trivial dynamics,
the current understanding of granular flows still remains
at its infancy.

Over the last two decades, the computer simulation studies on
granular shear flows have unveiled many interesting dynamical features
(such as cluster-formation, density-waves, plug formation,
stress-fluctuations, etc.) of such flows \cite{Luding1998,Luding2002}.
These studies have subsequently stimulated
many works on the associated macroscopic flows
described by suitable continuum equations. One major goal
of these theoretical works has been to test the available continuum models
from the viewpoint of predicting such pattern-formation in the rapid shear regime.
To this end, the kinetic-theory constitutive models \cite{Savage1992,Alam1998,Alam1998a}
have been routinely used.
The general concensus that appears to  have emerged 
is that the Navier-Stokes-level constitutive models
are able to explain many interesting dynamical states of
such rapid-shear flows.
We remark here that the non-Newtonian rheology is also important
for granular fluids \cite{Hutter1994,Alam2003}, 
but theoretical works with non-Newtonian 
constitutive models are still lacking.

In the present paper, we have performed 
linear instability and bifurcation analyses of the 
granular Couette flow under gravity using a 
kinetic-theory-based Newtonian hydrodynamic model.
Along with a synthesis of some recent work \cite{Alam2004},
here we provide new results on the effect of gravity on subcritical bifurcations
which firmly establish a link between our results
and the singularity theory of bifurcations.
We have shown that the origin of {\it shear-band} formation
in such flows is tied to an instability of the associated `gravity-free'
uniform shear flow via the route of the universal unfolding 
of pitchfork bifurcations.

\section{Hydrodynamic Model: Plane Couette Flow}

The hydrodynamic model that we have employed  is based on the kinetic theory
of inelastic hard-spheres \cite{Hutter1994}.
Under the assumption of instantaneous binary collisions,
the kinetic theory  provides a set of
hydrodynamic equations for the density $\tilde\rho=\tilde\rho_p\nu$,
the hydrodynamic velocity $\vec{\tilde u}$ and the granular energy $\tilde T$,
where $\nu$ is the volume fraction of particles
and  $\tilde \rho_p$ is their intrinsic density.
Note that the granular temperature is an additional ({\it higher-order}) field variable,
defined via $\tilde T=<\delta \vec{\tilde u}^2/3>$ where $\delta \vec{\tilde u}$ is the
fluctuation velocity  of a particle over the mean motion.
At the Navier-Stokes order, the hydrodynamic equations
\begin{subeqnarray}
 \left(\frac{\partial}{\partial t} + \vec{\tilde u}{\bf\cdot}{\bf\tilde\nabla}\right)
    \tilde\rho 
      &=& - \tilde\rho{\bf\nabla}{\bf\cdot}\vec{\tilde u}, \\
  \tilde\rho\left(\frac{\partial}{\partial t} 
    + \vec{\tilde u}{\bf\cdot}{\bf\tilde\nabla}\right)\vec{\tilde u} &=&
         - {\bf\tilde\nabla}{\bf\cdot}{\bf\tilde\Sigma} + \tilde\rho\vec{\tilde g}, \\
  \frac{3}{2}\tilde\rho\left(\frac{\partial}{\partial t} 
    + \vec{\tilde u}{\bf\cdot}{\bf\tilde\nabla}\right){\tilde T} &=&
        - {\bf\tilde\nabla}{\bf\cdot}\vec{\tilde q}
      - {\bf\tilde\Sigma}{\bf :}{\bf\tilde\nabla}\vec{\tilde u} - \tilde{\mathcal D},
\end{subeqnarray}
represent the mass, momentum and energy balances, respectively.
Here ${\bf\tilde\Sigma}$ is the stress tensor,
$\vec{\tilde q}$  the energy-flux,
${\bf\tilde\Sigma}{\bf :}{\bf\tilde\nabla}\vec{\tilde u}$ the production
of fluctuation energy due to shear-work,  $\tilde{\mathcal D}$ the
rate of collisional dissipation of fluctuation energy
and $\vec{\tilde g}$ is the acceleration due to gravity.
Note that the collisional dissipation rate vanishes identically
for perfectly elastic collisions (i.e. the coefficient of restitution is $e=1$). 
As mentioned, the third equation is a
balance for the fluctuation kinetic energy
which is needed since the transport properties of granular fluids depend on
the granular temperature.
We employ the standard Newtonian constitutive model of
an inelastic dense gas
for the stress tensor ${\bf\tilde\Sigma}$,
the energy flux $\vec{\tilde q}$ and the collisional dissipation rate
$\tilde{\mathcal D}$ as detailed in \cite{Lun1984}.

\begin{figure}[htb]
\begin{center}
\includegraphics[width=.7\textwidth]{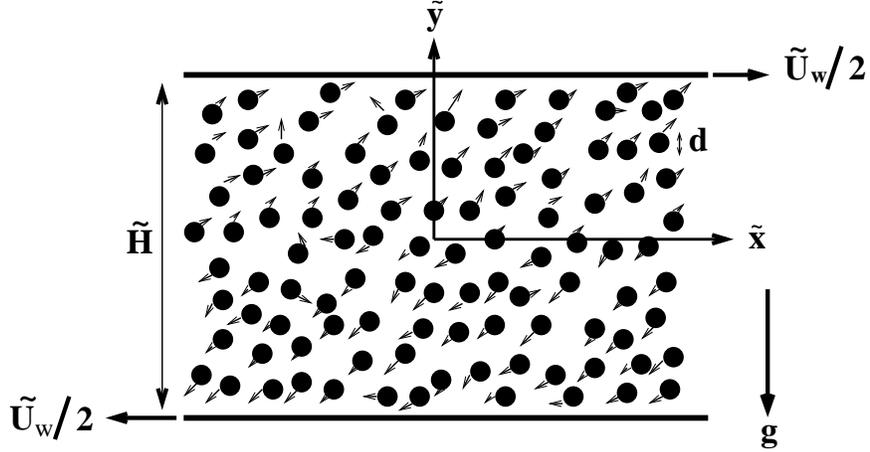}
\end{center}
\caption[]
{
A schematic of the plane Couette flow of monodisperse granular materials,
with the particle size being $\tilde{d}$
}
\label{fig:fig1}
\end{figure}

The plane Couette flow under gravity, as shown schematically in Fig. 1,
is driven by two oppositely-moving
walls at $\tilde y = - {\tilde H}/2$ and $\tilde y = {\tilde H}/2$.
In the Cartesian coordinate system,
$\tilde{u}(x,y,t)$ and $\tilde{v}(x,y,t)$ denote the
streamwise and transverse velocity components in
the $\tilde x$- and $\tilde y$-directions, respectively.
Using the separation between the two walls $\tilde H$ as the 
length scale, the velocity difference
between the walls $\tilde{U}_w$ as the velocity scale and the inverse of the overall shear
rate ${\tilde H}/{\tilde U}_w=\tilde{\gamma}^{-1}$ as the time scale,
one can obtain the non-dimensional equations of motion in terms of
dynamical variables $\nu$, $\vec{u}= (u, v)$ and $T$ \cite{Alam2004}.
With this non-dimensionalization procedure, 
the top and bottom walls move with velocities $\pm 1/2$, respectively.

For the steady ($\partial/\partial t=0$),
fully developed ($\partial/\partial x=0$) plane Couette flow
under gravity, the mass balance equation is identically satisfied.
The  $x$-momentum, $y$-momentum and energy balance
equations reduce to
\begin{subeqnarray}
  \frac{\rm d}{{\rm d}y}\left(\mu\,
     \frac{{\rm d}u}{{\rm d}y}\right) &=& 0 , 
\label{eqn_PCF1} \\
 \frac{{\rm d}p}{{\rm d}y} + \frac{\nu H^3}{Fr^2}  &=& 0 , 
\label{eqn_PCF2} \\
 H^{-2}\frac{{\rm d}}{{\rm d}y}\left(\kappa\,
     \frac{{\rm d}T}{{\rm d}y}
     \right)
     + \mu\,\left(\frac{{\rm d}u}{{\rm d}y}\right)^2 - {\mathcal D} &=& 0, 
\label{eqn_PCF3}
\end{subeqnarray}
respectively.
Here $p(\nu,T)=f_1(\nu,e)T$, $\mu(\nu,T)=f_2(\nu,e)\sqrt{T}$,
$\kappa(\nu,T)=f_4(\nu,e)\sqrt{T}$
and  ${\mathcal D}(\nu,T)=f_5(\nu,e)T^{3/2}$
are the nondimensional forms of the pressure, shear viscosity,
pseudo-thermal conductivity and the collisional dissipations,
respectively.
The explicit forms for these non-dimensional
functions $f_1-f_5$ are detailed elsewhere \cite{Alam1998}.

Only four non-dimensional control parameters are needed to describe
the present problem: 
the average solid fraction $\nu_{av}$,
the non-dimensional Couette gap $H=\tilde{H}/\tilde{d}$,
the coefficient of restitution $e$, 
and the Froude number $Fr=\tilde{U}_w/\sqrt{\tilde{g}\tilde{d}}$
(or, the non-dimensional weighted shear rate).
Note that $Fr=\gamma\tau_d$ can be written as a ratio between two time-scales,
where $\gamma=\tilde{U}_w/\tilde{H}$ is the external shear rate
and $\tau_d=H(\tilde{d}/\tilde{g})^{1/2}$ is the gravitational time scale.
The case of zero-gravity (i.e. $g=0$)
corresponds to $Fr^{-1}=0$ which can also be recovered
by considering the infinite shear-rate limit ($\gamma\to\infty$).

For all our calculations, we have adopted the
{\it no-slip} and {\it zero} energy-flux boundary conditions.
(The effects of slip-velocity and non-zero energy-flux 
do not introduce any new physics \cite{Alam1998,Nott1999,Alam2004}  since the reported
shear-band formation is driven by a bulk-instability.)  
With these ideal boundary conditions, the gravity-free case admits
a uniform shear solution with constant solid fraction and granular energy:
\begin{equation}
  \nu(y) = \mbox{const. }, \quad
  u(y) = y, \quad
  T(y) = f_2(\nu, e)/f_5(\nu,e).
\label{eqn_USF}
\end{equation}
It is easy to verify that equations (\ref{eqn_PCF1}--\ref{eqn_PCF3})
with $Fr^{-1}=0$ admit the following symmetry:
\begin{equation}
  \nu(y) = \nu(-y), \quad
  u(y) = - u(-y), \quad
  T(y) = T(-y).
\label{eqn_SYM}
\end{equation}
i.e. the solid fraction and granular temperature are symmetric
about the mid-plane ($y=0$), and the velocity is antisymmetric
for the uniform shear flow.
However, for the plane Couette flow with gravity ($Fr^{-1}\neq 0$),
this {\it center-symmetry} is no longer preserved.

\section{Shear-banding Instability: Sub- and Supercritical Pitchfork Bifurcations}

For instability analysis, the uniform shear  flow is
perturbed by infinitesimal perturbations, viz.,
\begin{subeqnarray}
   \nu(x,y,t) &=& \nu + \nu'(x,y,t) , \\
   u(x,y,t) &=& y + u'(x,y,t) , \\
   v(x,y,t) &=&  v'(x,y,t) , \\
   T(x,y,t) &=&  f_2/f_5 + T'(x,y,t) .
\end{subeqnarray}
The dynamics of perturbations  is then studied by linearizing the
equations of motion around the base flow $(\nu, y, 0, f_2/f_5)$
via the normal-mode analysis,
\begin{equation}
{\left[\nu', u', v', T'\right](x,y,t)} =
    {[\hat\nu, \hat u, \hat v, \hat T](y)\,{\E}^{{\I} k_x x + \omega t} } .
\end{equation}
Here $k_x$ is the wavenumber for the $x$-direction and  
the quantities with hats are complex amplitudes of perturbations;
$\omega=\omega_r + i\omega_i$ is the complex frequency.
Instability is associated with positive values of $\omega_r$.

It is known that the uniform granular shear flow is unstable to
various kinds of disturbances, leading to both
stationary and travelling wave patterns \cite{Alam1998}.
Here we focus only on a special kind of {\it stationary} instability that
arises due to purely streamwise disturbances ($k_x=0$),
i.e. the disturbance patterns do not vary with $x$;
we call it {\it shear-banding instability} since its nonlinear
saturation leads to shear-banding-type patterns of alternating bands of
dilute and dense regions in the gradient direction.
For such disturbances,  there is
a minimum value of solid fraction ($\nu_{av} \sim 0.15$)
above which the flow is unstable if the Couette gap 
satisfies the following relation:
\begin{equation}
   H \geq  n\pi \psi(\nu,e), 
\end{equation}
where $\psi(\nu,e)$ is a complicated function of density and restitution coefficient,
and $n=1,2,\ldots$ is the mode number (which is related
to the eigenfunctions of the linearized stability problem).

It is clear that the $n=1$ mode is the first to become unstable
at a critical value of the Couette gap $H=H_c\equiv \pi \psi(\nu,e)$ 
(for given $\nu_{av}$ and $e$).
Beyond $H>H_c$, the successive higher-order modes ($n=2,3,\ldots$) take over
as the most unstable mode at $H=nH_c$.

Treating $H$ as a bifurcation parameter, there is now
a countably infinite number of pitchfork bifurcations 
(since the least-stable eigenvalue is real), located at $H=nH_c$.
Before presenting bifurcation results, let us briefly introduce 
the concept of pitchfork bifurcations by
considering the simplest bifurcation 
problem with one {\it order} parameter $\Phi$,
\begin{equation}
  \frac{{\rm d}\Phi}{{\rm d}t} = f(\Phi, {\mathcal B})
       = b\Phi^3 - {\mathcal B}\Phi = -  f(-\Phi, {\mathcal B}),
\label{eqn_nform}
\end{equation}
where $f(\Phi, {\mathcal B})$ is a nonlinear scalar valued function of $\Phi$,
${\mathcal B}$,  called the {\it bifurcation} ({\it control})  parameter,
and $b$ is some real parameter.
The bifurcation diagram consists of the set of points in
the $(\Phi, {\mathcal B})$-plane
which  satisfies $f(\Phi, {\mathcal B})=0$.
For pitchfork (stationary) bifurcations,
the number of solutions
of $f(\Phi, {\mathcal B})=0$ jumps from one to three
as the bifurcation
parameter ${\mathcal B}$ crosses some critical value ${\mathcal B}_c$.
More specifically, the loss of stability of the base solution $\Phi=0$
gives birth to two {\it new} solutions,
\[
   \Phi = \pm\sqrt{{\mathcal B}/b} , 
\]
which are {\it stable}. These solutions,
$\Phi = \pm\sqrt{{\mathcal B}/b}$, are said to be {\it bifurcated}
from the base solution $\Phi=0$.
The bifurcation is called {\it supercritical} for $b>0$,
and {\it subcritical} for $b<0$.

\begin{figure}[htb]
\begin{center}
\includegraphics[width=.35\textwidth]{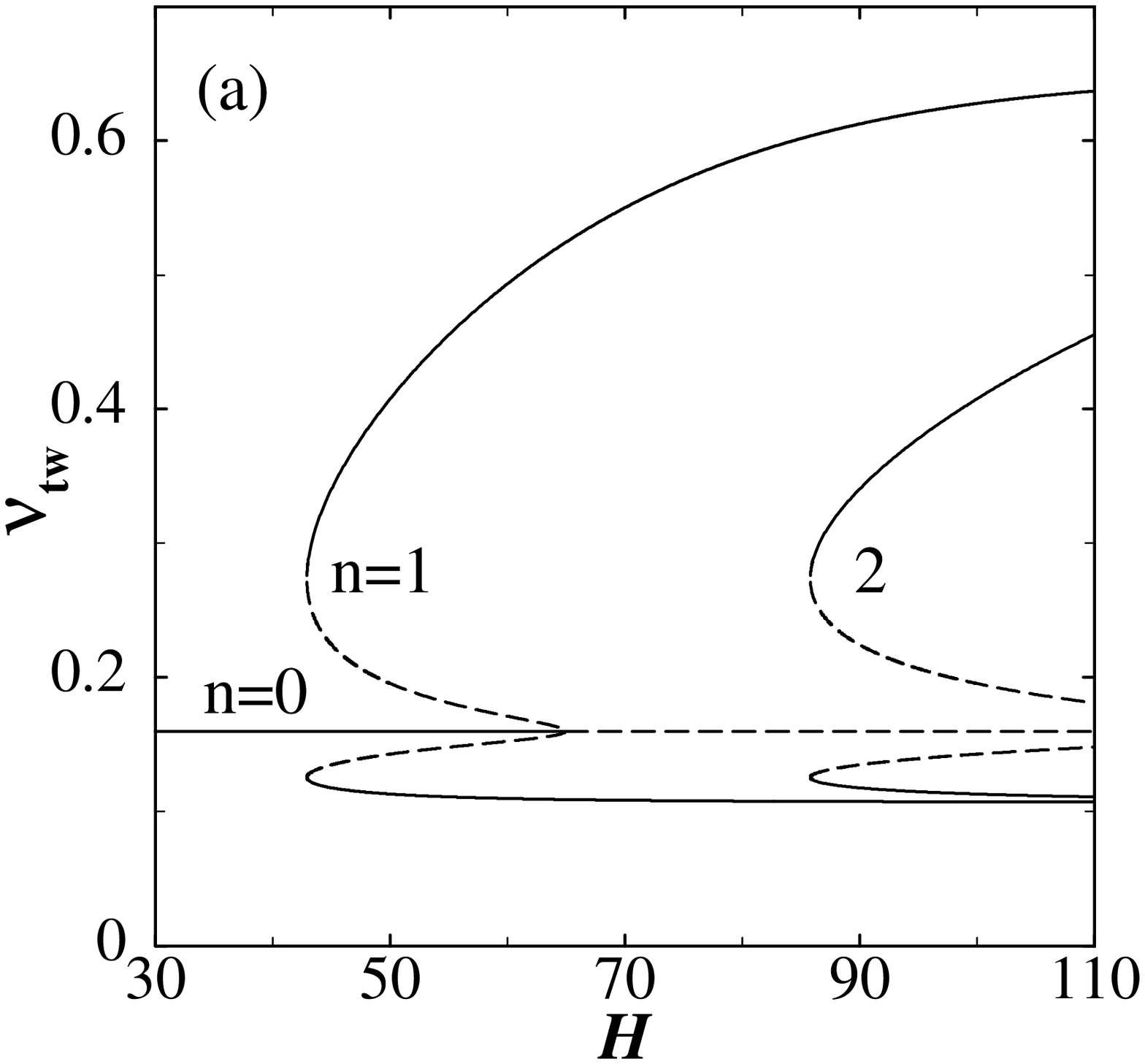}
\includegraphics[width=.35\textwidth]{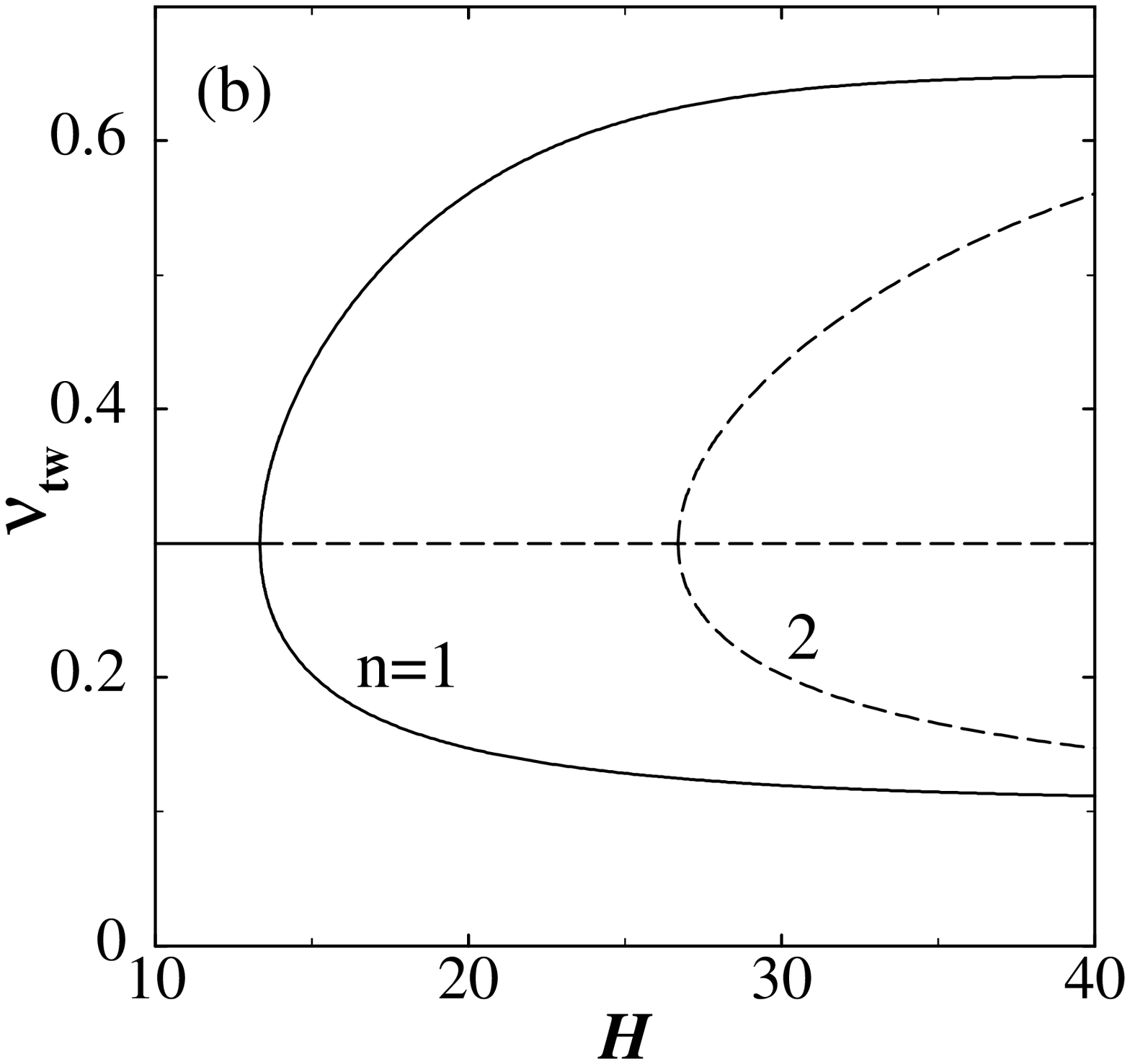}\\
\includegraphics[width=.35\textwidth]{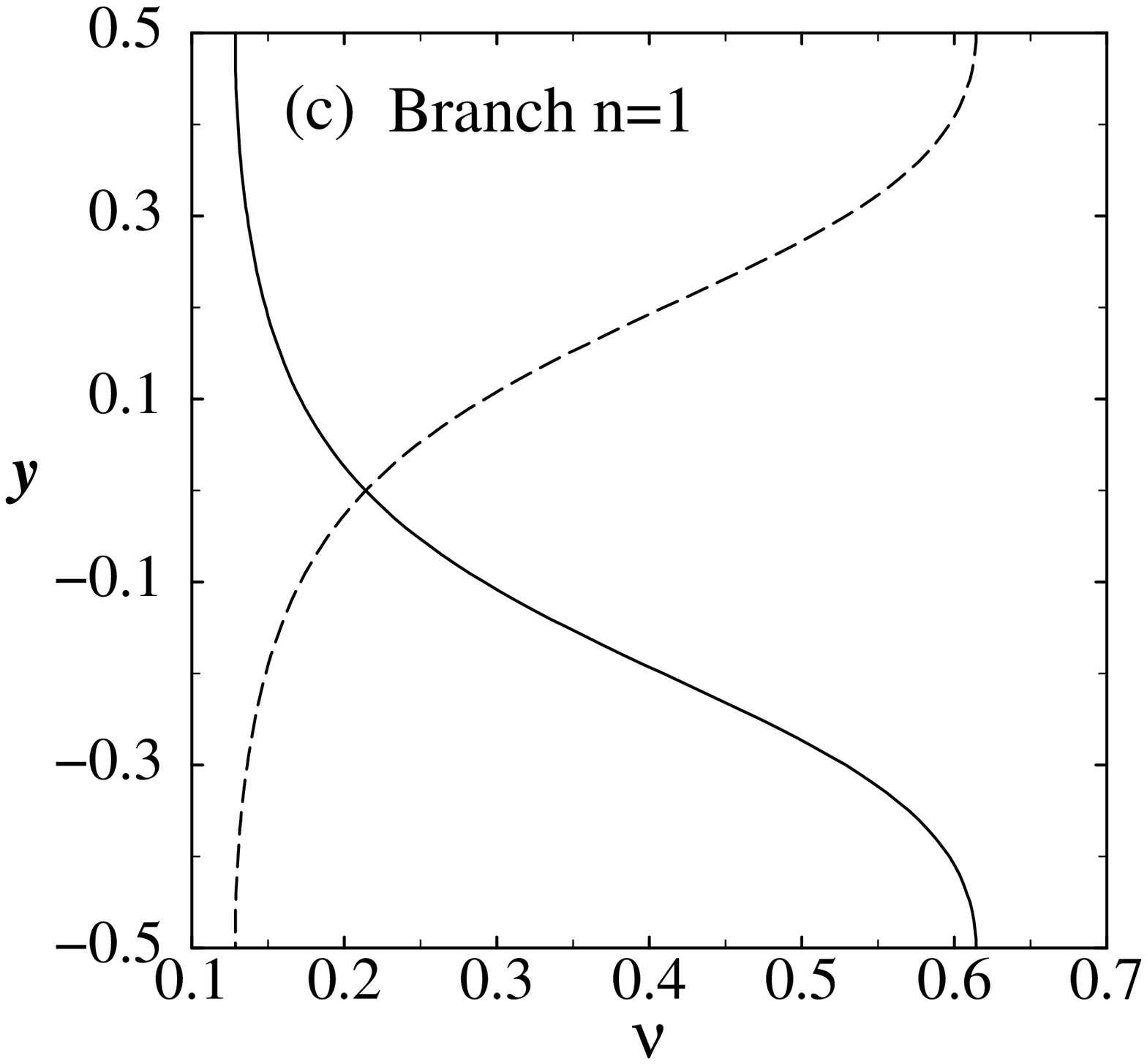}
\includegraphics[width=.35\textwidth]{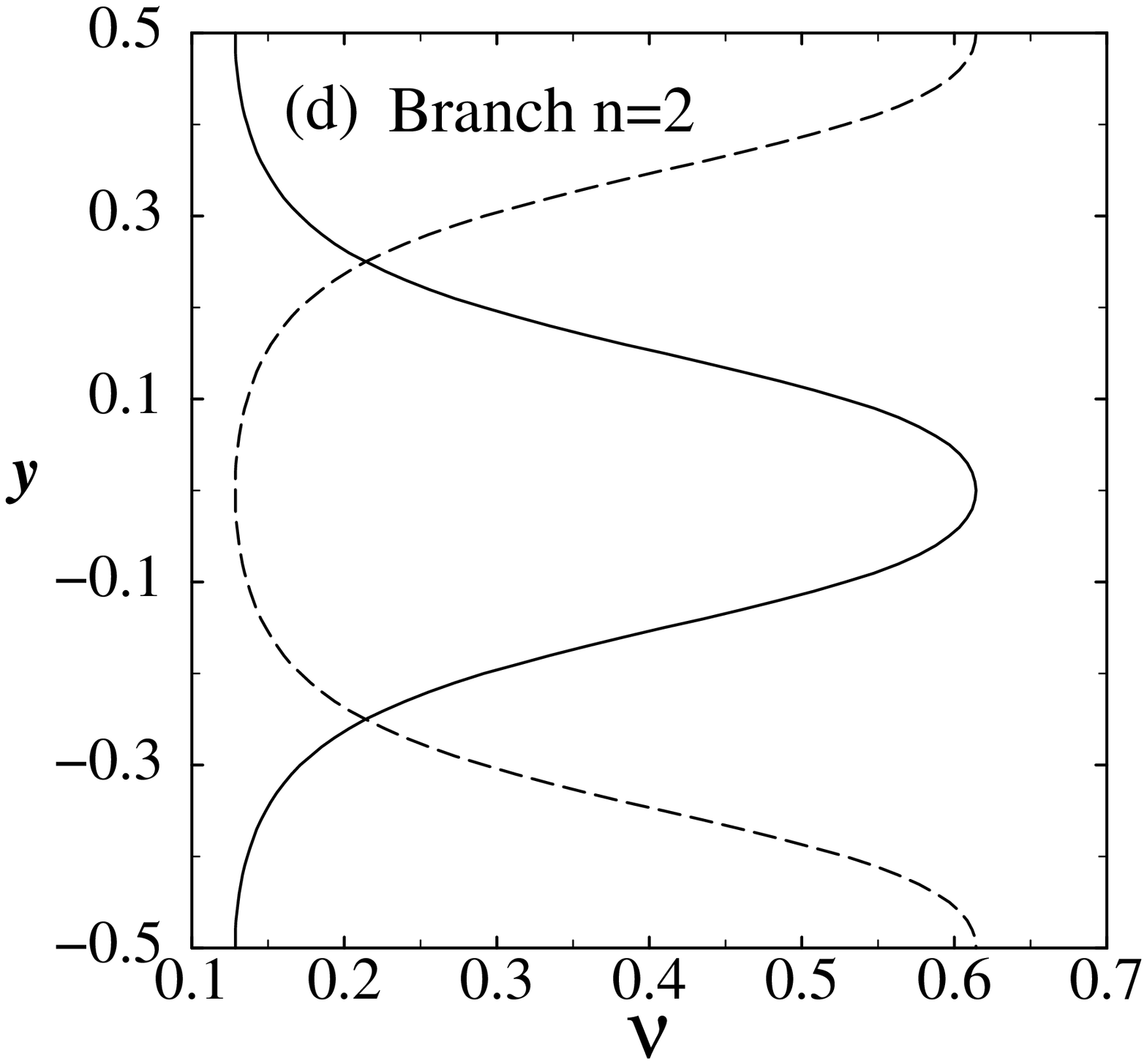}
\end{center}
\caption[]
{
Evidence of subcritical and supercritical bifurcations
in granular Couette flow: 
the average solid fraction $\nu_{av}$ is (${\bf a}$) $0.16$,
(${\bf b}$) $0.3$;
the coefficient of restitution is $e=0.8$.  
The {\it stable} (/{\it unstable}) bifurcation
branch is denoted by solid (dashed) line.
(${\bf c}$-${\bf d}$) Variations of solid fraction
for (${\bf c}$) the first-pair ($n=1$) of solution branches at $H=25$
and for (${\bf d}$) the second-pair ($n=2$) of solution branches at $H=50$
with $\nu_{av}=0.3$;
in each panel, the solid and dashed lines represent solutions
corresponding to 
lower and upper bifurcation branches in panel (${\bf b}$), respectively
}
\label{fig:fig2}
\end{figure}

For the present problem,
we have chosen the {\it order} parameter to be
$\Phi=\nu(1/2)=\nu_{tw}$, the value of solid fraction at the top-wall,
(this can, equivalently, be tied to $(\nu_{tw}-\nu_{av})$ or
any other suitable quantity),
and the {\it bifurcation} parameter as
${\mathcal B}=H$ (this can also be tied to $(H-H_c)$,
i.e. ${\mathcal B}\equiv {\mathcal B}(H,e,\nu_{av})$).
Figure \ref{fig:fig2}($a$) shows a typical subcritical bifurcation diagram 
for the uniform shear case in the ($H, \nu_{tw}$)-plane
for a low density granular fluid ($\nu_{av}=0.16$);
the coefficient of restitution is  $e=0.8$.
The horizontal line refers to uniform shear
for which the density is uniform across the Couette gap,
i.e.  $\nu_{tw}\equiv \nu_{av}$.
As we increase the bifurcation parameter $H$, the
uniform shear flow loses stability to the first mode ($n=1$)
at $H=H_c\approx 64.4$ and two {\it unstable} solution branches emerge
from this bifurcation point.
Each of these unstable branches  eventually
turns over at some lower value of $H=H_u\approx 42$ to yield 
a {\it stable} solution branch.
For the range of Couette gaps $H_u< H< H_c$, there
are three possible solutions, with the middle solution
branch being unstable.
This is an example of {\it hysteretic} phase transition.

As we increase the average density, the nature of bifurcation
changes from {\it subcritical} to  {\it supercritical} as seen in
Fig. \ref{fig:fig2}($b$)
for a moderate density granular fluid ($\nu_{av}=0.3$).
For these parameter combinations, the bifurcation
is subcritical for $\nu_{av} < 0.18$ and supercritical for $\nu_{av} > 0.18$.
From each bifurcation point, two stable solution branches emerge and
such pairs of solutions arise due to the underlying symmetry of the
uniform shear flow.
Note that all the upper bifurcation branches asymptote
to $\nu_{tw}\to 0.65 = \nu_{max}$ which corresponds to
the maximum allowable solid fraction in the system,
representing the random close-packing limit.

For the sake of completeness, we show the density profiles
of the first two branches in Fig. \ref{fig:fig2}($b$)
at $H=25$ and $50$, respectively, in Figs. \ref{fig:fig2}($c$-$d$).
The solid and dashed lines in Figs. \ref{fig:fig2}($c$-$d$) 
represent solutions belonging to the lower and upper branches
of Fig. \ref{fig:fig2}($b$), respectively.
The solutions emerging from $H=H_c, 3H_c,\ldots$ are {\it asymmetric},
having no symmetry about the centerline ($y=0$),
since the eigenfunctions for odd modes ($n=1, 3, \ldots$)
destroy the center-symmetry (\ref{eqn_SYM}) of uniform shear.
On the other hand, the solutions from $H=2H_c, 4H_c,\ldots$ 
are {\it symmetric} about the centerline,
since the eigenfunctions for even modes ($n=2, 4, \ldots$) satisfy 
the symmetry (\ref{eqn_SYM}).
Related bifurcation results on the gravity-free uniform shear flow 
with multiple plugs are detailed elsewhere \cite{Nott1999}.
In the following, we consider the role of gravity on the
fate of such plugged solutions.

\begin{figure}[htb]
\begin{center}
\includegraphics[width=.35\textwidth]{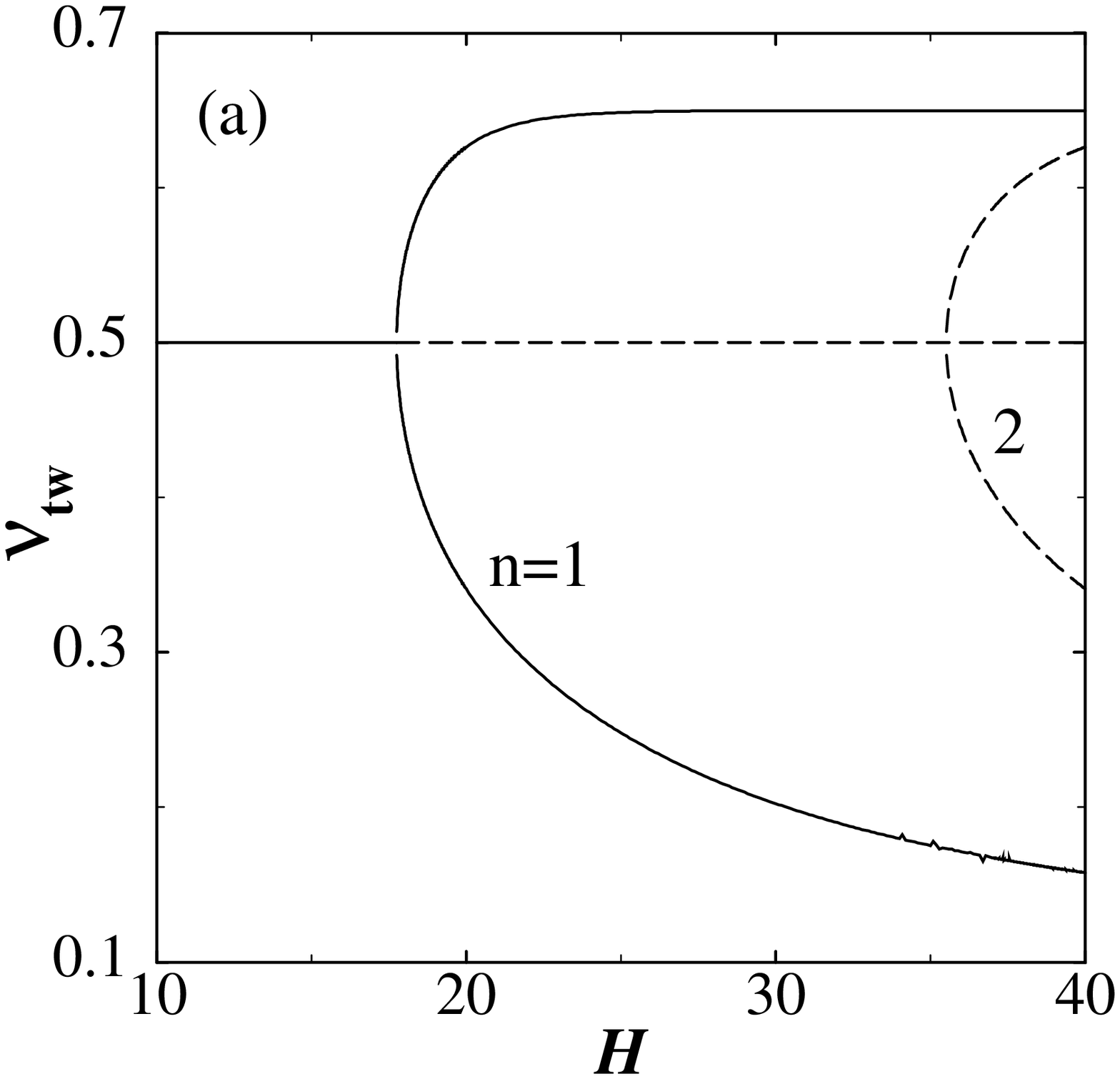}
\includegraphics[width=.35\textwidth]{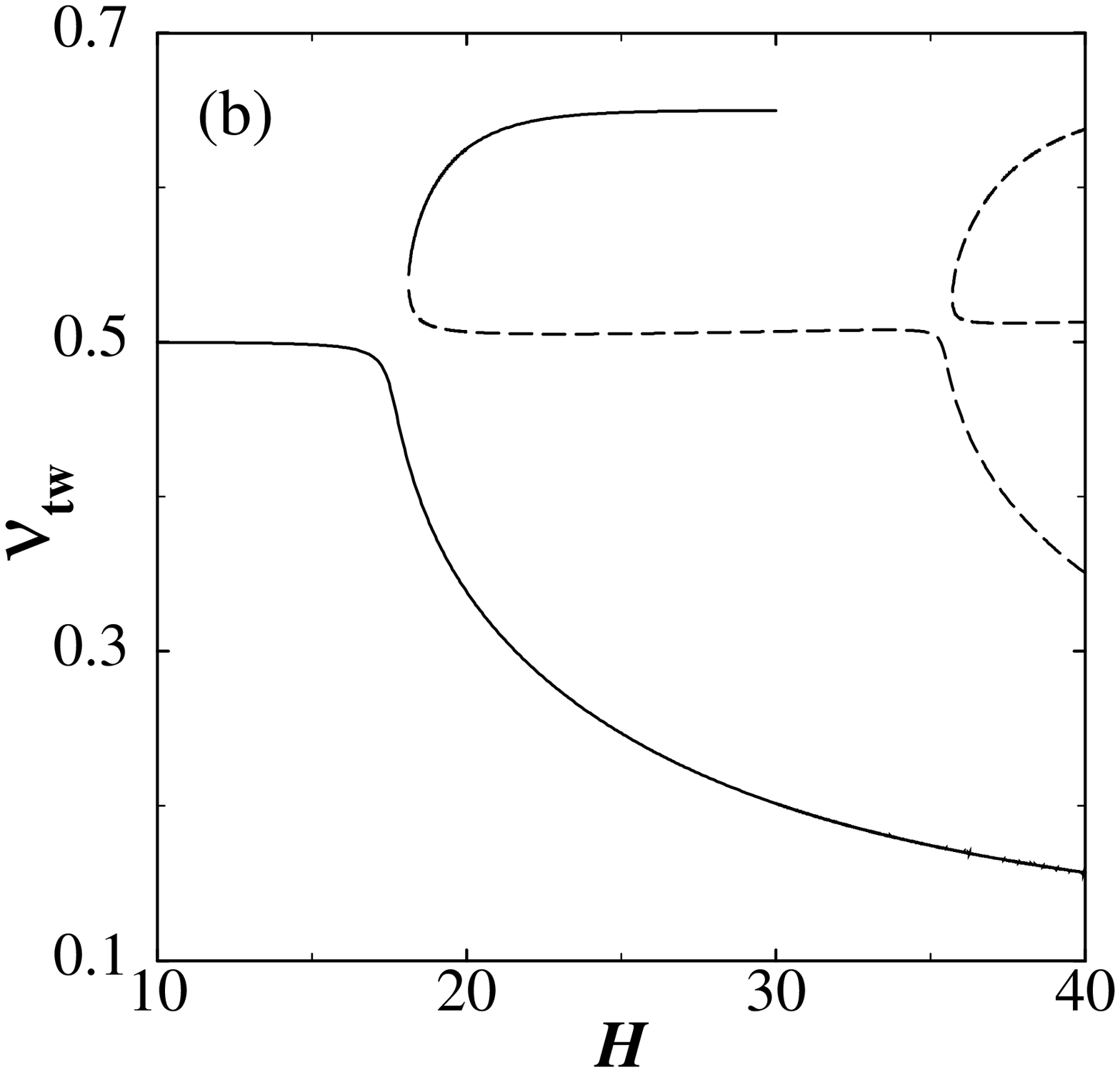}\\
\includegraphics[width=.35\textwidth]{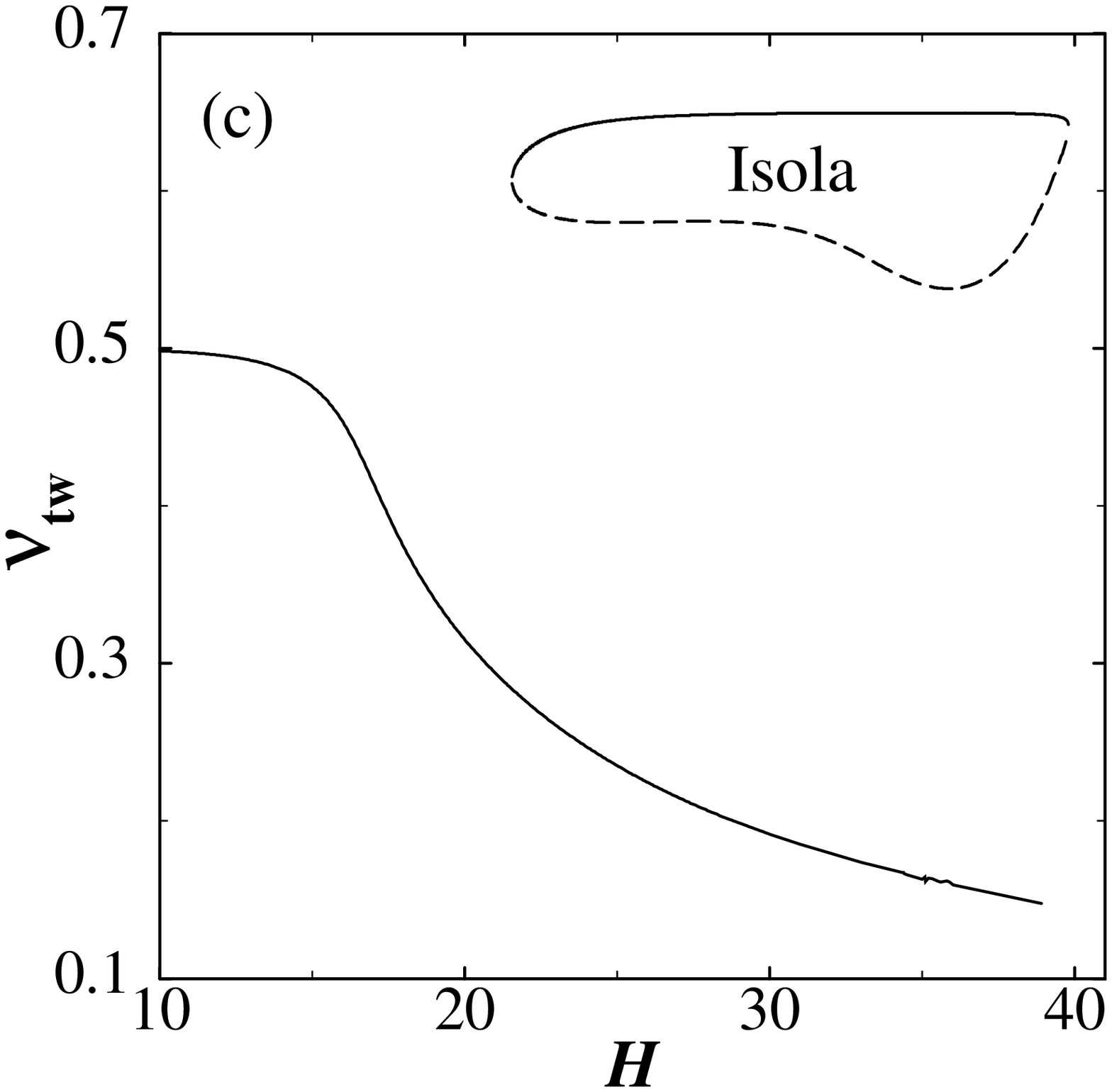}
\end{center}
\caption[]
{
Effect of gravity on the bifurcation diagram
with $\nu_{av}=0.5$ and $e=0.8$:
(${\bf a}$) $Fr^{-1} = 0$;
(${\bf b}$) $Fr^{-1} =  10^{-3}$;
(${\bf c}$) $Fr^{-1} = 4\times 10^{-3}$.
The {\it stable} ({\it unstable})
bifurcation branch is denoted by solid (dashed) line
}
\label{fig:fig3}
\end{figure}

\section{Gravitational Plane Couette Flow: Gravity as an Imperfection}

\begin{figure}[htb]
\begin{center}
\includegraphics[width=.35\textwidth]{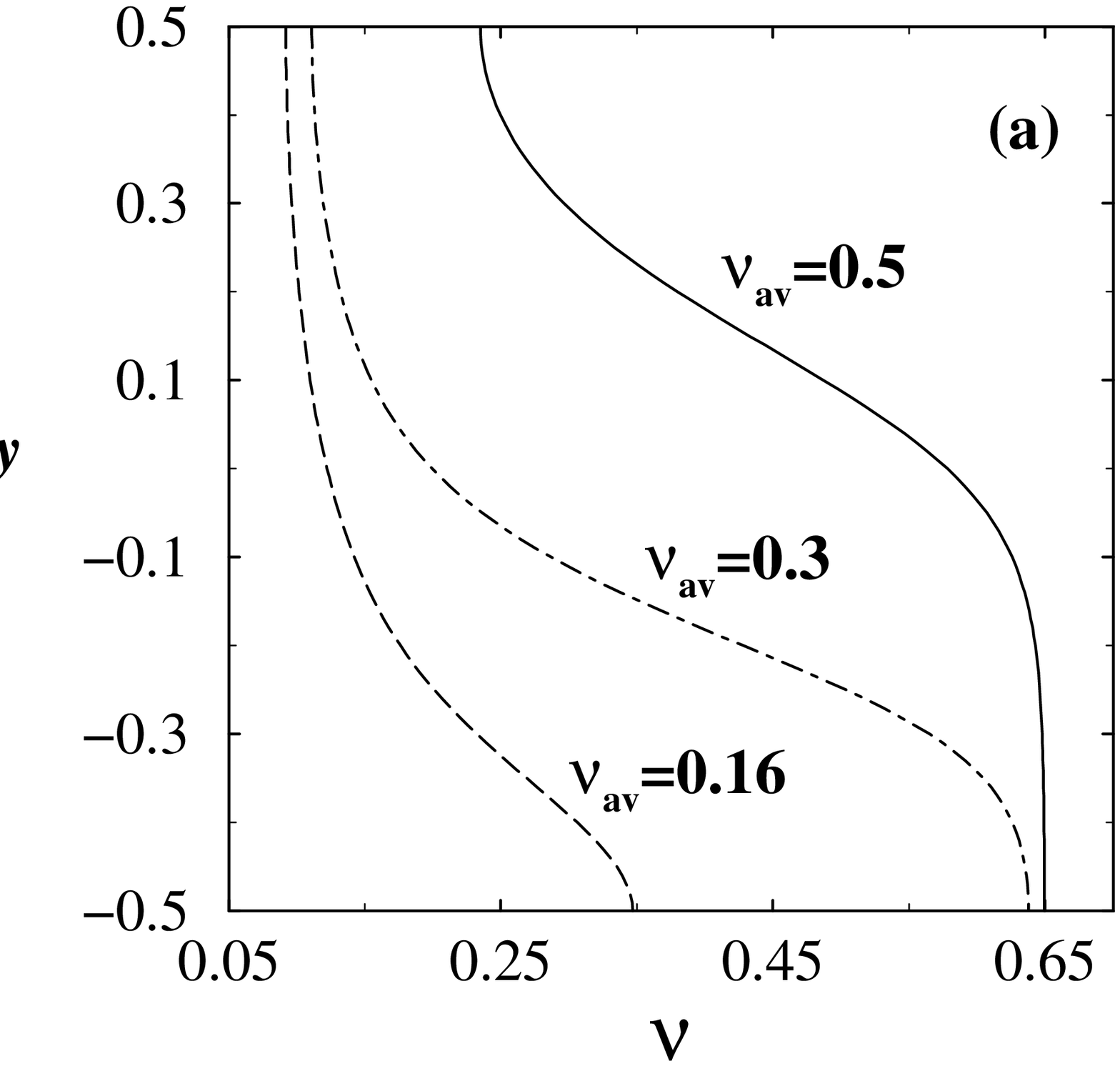}
\includegraphics[width=.35\textwidth]{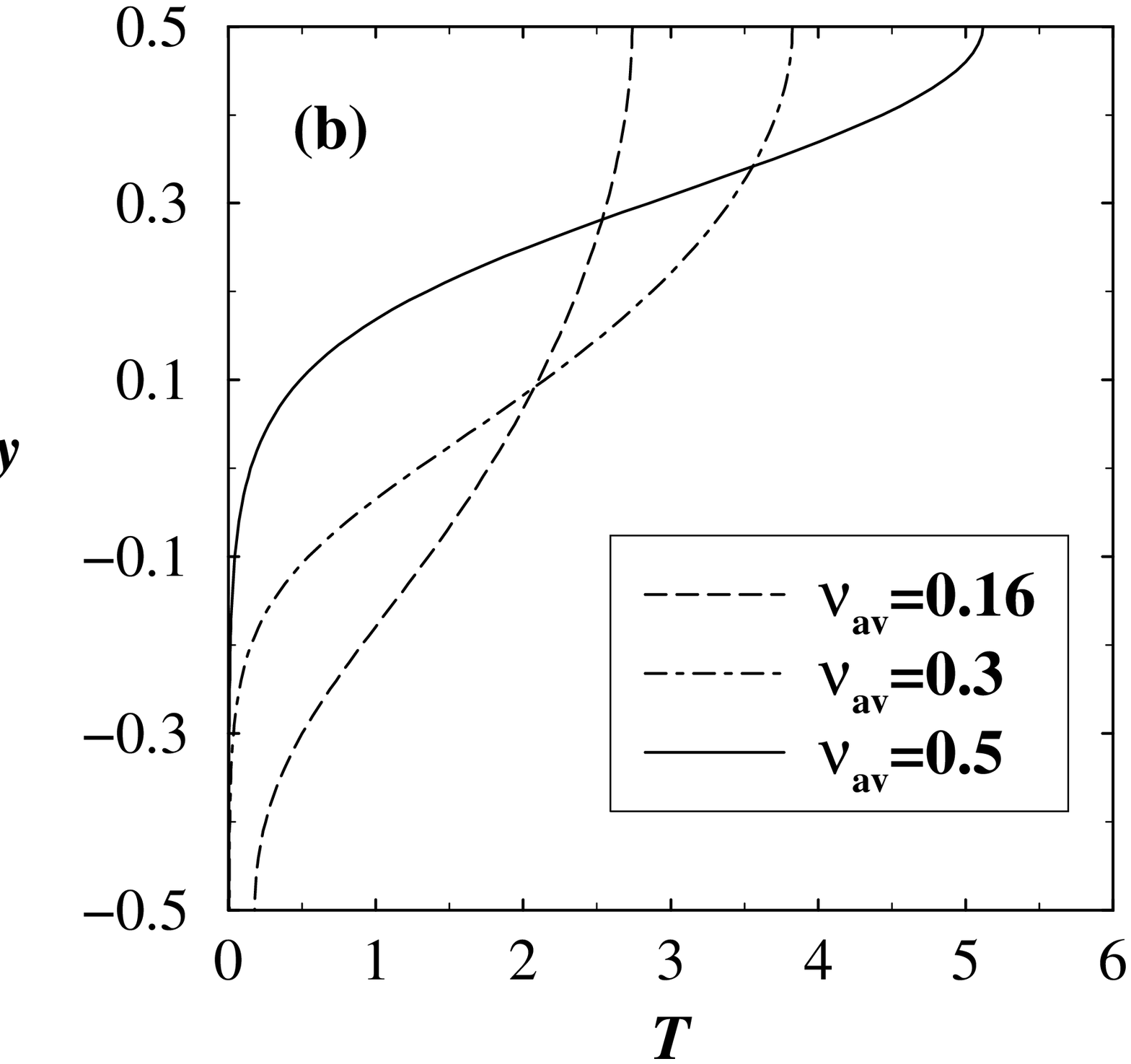}\\
\includegraphics[width=.35\textwidth]{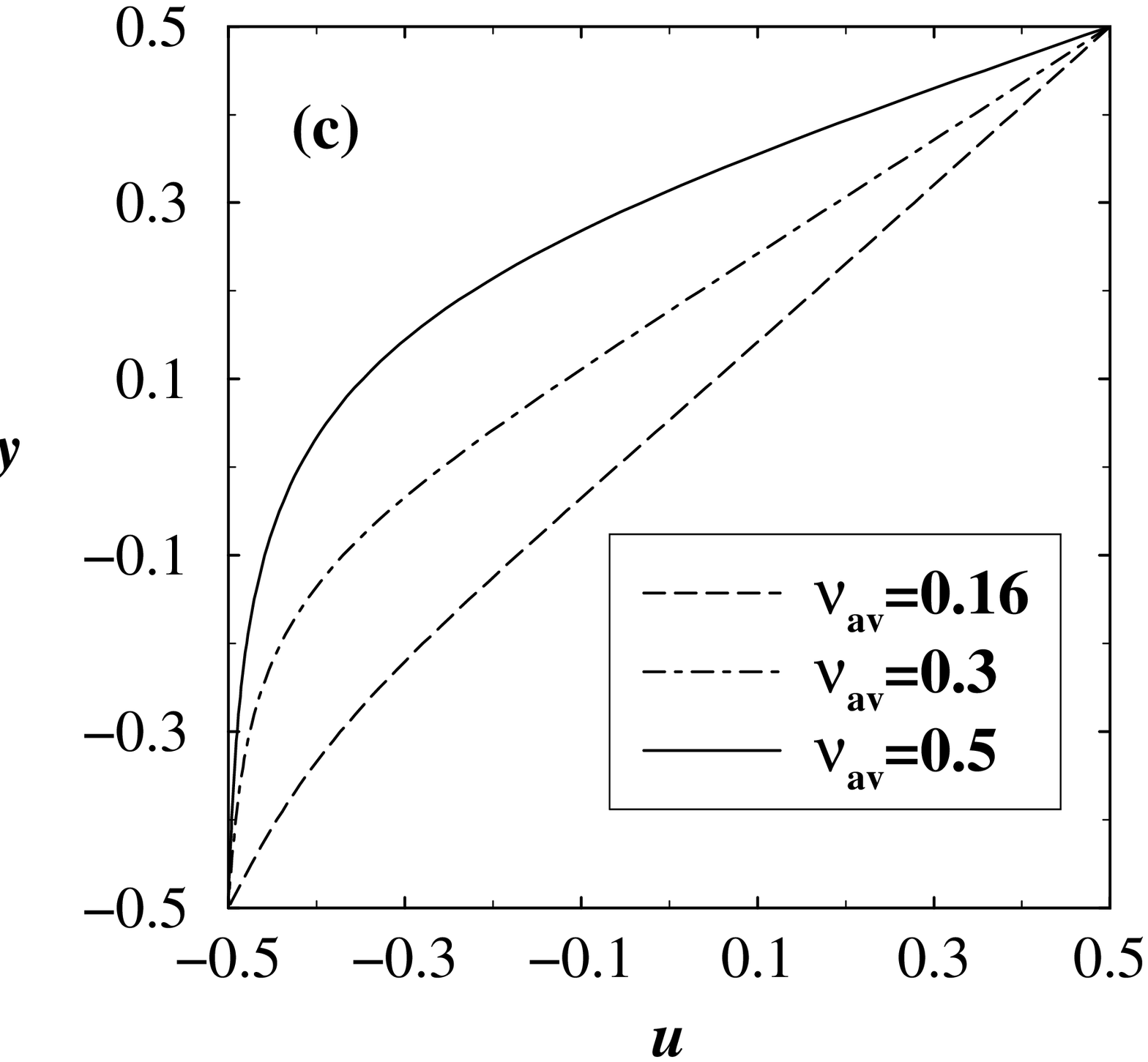}
\end{center}
\caption[]
{
Effect of average density, $\nu_{av}$, on the profiles
of (${\bf a}$) volume fraction, (${\bf b}$) granular temperature
and (${\bf c}$) streamwise velocity
at $Fr^{-1}= 4\times 10^{-3}$, $H=25$ and $e=0.8$. 
The solutions correspond to the 
`unique' lower bifurcation branch
}
\label{fig:fig4}
\end{figure}

Figures \ref{fig:fig3}(${\bf a}$-${\bf c}$)
display three bifurcation diagrams at different $Fr^{-1}$,
showing the effect of gravity on the topology of bifurcations.
The average density is set to $\nu_{av}=0.5$,
representing dense flows, and other parameters
are as in Fig. \ref{fig:fig2}.
For the gravity-free case ($Fr^{-1}=0$) in Fig. \ref{fig:fig3}(${\bf a}$),
the topology of bifurcations is similar to that in Fig. \ref{fig:fig2}($b$).
As we increase the gravitational strength 
to $Fr^{-1}=10^{-3}$ (Fig. \ref{fig:fig3}$b$), the
solution branches from the even and odd modes peels off from each 
bifurcation point;
this is an example of {\it imperfect} bifurcation.
Thus,  {\it gravity acts as an imperfection in the embedded problem},
destroying the bifurcation structure of the gravity-free case.
By further increasing the gravitational strength to $Fr^{-1}= 4\times 10^{-3}$ 
(Fig. \ref{fig:fig3}$c$), the
solution branches are seen to form {\it isolas}, via
merging between two successive bifurcation branches.
Interestingly, the size of this isola also decreases
with increasing $Fr^{-1}$ which disappears completely at $Fr^{-1}\sim 5\times 10^{-3}$.
This simply implies that  gravity does not support the associated solutions.
(The sequence of {\it birth} and {\it death} of such isolas 
is detailed elsewhere \cite{Alam2004}.)
The point we want to emphasize is that
there is  a critical value of $Fr$ ($\sim 5\times 10^{-3}$ for
the parameter combinations of Fig. \ref{fig:fig3})
below which only
{\it one}  solution-branch survives in the bifurcation diagram
that corresponds to the lower branch of the first mode ($n=1$)
which extends upto $H=\infty$.
Note that the density profile corresponding to this
`unique' branch has a plug near the bottom plate as shown by the solid line in
Fig. \ref{fig:fig4}$a$.

Figures \ref{fig:fig4}($a$-$c$) show the effect of mean density on the
profiles of volume fraction, granular temperature and  velocity
at $Fr^{-1}= 4\times 10^{-3}$, $H=25$ and $e=0.8$. 
The dashed, dot-dashed and solid lines
represent solutions at $\nu_{av}=0.16$, $0.3$ and $0.5$, respectively.
We observe that the thickness of the plug increases with increasing $\nu_{av}$
and the plug becomes compactified.
Within the plug, both the granular temperature and the shear-rate
are relatively small.
With increasing mean density, the plug becomes almost {unsheared}
(i.e. the local shear rate being close to zero)
near the bottom plate.

\subsection{Connection with Universal Unfolding of Pitchfork Bifurcation}

We have established that gravity acts as an imperfection in the present problem.
In the parlance of {\it singularity} theory \cite{GS1985},
the general problem of imperfect bifurcation
is addressed by using the theory of `universal' unfolding which
we discuss in this section.

\begin{figure}[htb]
\begin{center}
\includegraphics[width=.6\textwidth]{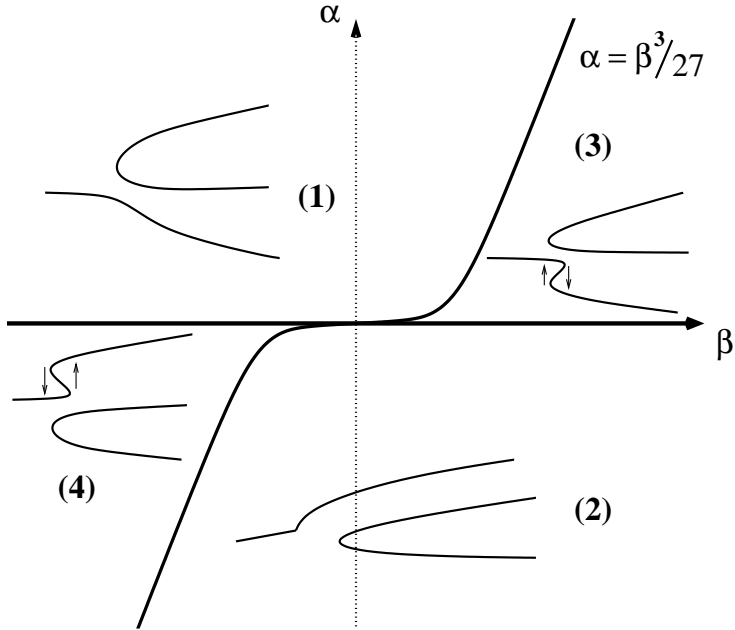}
\end{center}
\caption[]
{
A schematic phase-diagram for the universal unfolding of pitchfork bifurcations,
showing all possible forms of imperfect bifurcations
}
\label{fig:fig5}
\end{figure}

To connect our bifurcation results with {\it universal unfoldings},
let us consider the {\it normal-form} equation 
for {\it imperfect}  pitchfork bifurcation \cite{GS1985},
\begin{equation}
  \frac{{\rm d}\Phi}{{\rm d}t} = \Phi^3 - {\mathcal B}\Phi + \alpha + \beta \Phi^2
              = F(\Phi, {\mathcal B}; \alpha, \beta),
\label{eqn_nform1}
\end{equation}
where $\alpha$ and $\beta$ are the imperfection parameters
which vanish identically for `perfect' pitchfork bifurcations
such that
\begin{equation}
  F(\Phi, {\mathcal B}; 0, 0) \equiv f(\Phi, {\mathcal B}) .
\end{equation}
From the equations of motions,
we can assert  that these imperfection parameters $\alpha$ and $\beta$ are functions
of $Fr$, $H$ and $\nu$,
i.e.
\begin{subeqnarray}
 \alpha &\equiv& \alpha(Fr^{-1}, H, \nu)\\
 \beta &\equiv & \beta(Fr^{-1}, H, \nu) ,
\end{subeqnarray}
satisfying the following properties:
\begin{equation}
   \alpha(0, H, \nu) = 0 = \beta(0, H, \nu) .
\end{equation}
In fact, one can think of $F(\Phi, {\mathcal B}; \alpha, \beta)$ 
as a perturbation of the scalar valued function $f(\Phi, {\mathcal B})$.
This function $F(\Phi, {\mathcal B}; \alpha, \beta)$ is 
called the universal unfolding of $f(\Phi, {\mathcal B})$
since it encompasses all possible forms of 
bifurcation diagrams that can arise due to any form
of perturbations on the perfect pitchfork bifurcation.

As discussed in  \cite{GS1985},
the universal unfolding of the pitchfork leads to four
possible bifurcation diagrams in the $(\alpha, \beta)$-plane.
In particular, two curves $\alpha=0$ and $\alpha=\beta^3/27$,
denoted by thick solid lines in Fig. \ref{fig:fig5},
divide the $(\alpha, \beta)$-plane into four
zones, and all parameter combinations in each zone
represent a canonical bifurcation diagram as shown schematically in this figure.
Comparing our bifurcation diagrams
in Fig. \ref{fig:fig3} 
with these four sets, we find that the bifurcation diagram for zone $1$
represents the unfolding of pitchfork bifurcations in our case.
However, by redefining our order-parameter as $\Phi=\nu_{bw}\equiv \nu(-1/2)$,
i.e. the value of the solid fraction at the bottom wall, and then redrawing
the bifurcation diagrams of Fig. \ref{fig:fig3},
we  get back the canonical set for zone $2$.
For this case,  the smooth solution branch
appears from the upper part of the corresponding $n=1$ bifurcation branch
of the gravity-free case.

\begin{figure}[htb]
\begin{center}
\includegraphics[width=.35\textwidth]{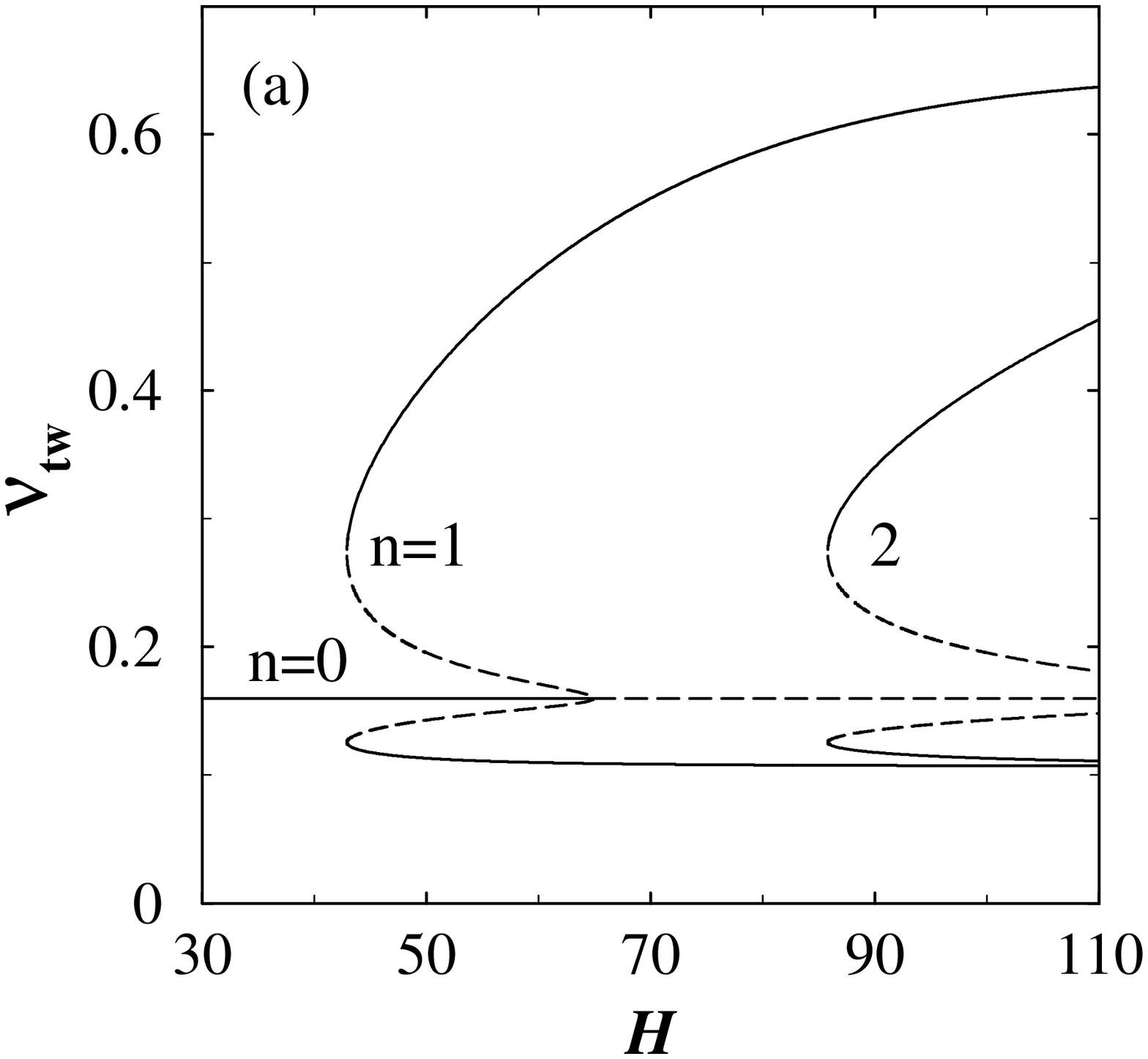}
\includegraphics[width=.35\textwidth]{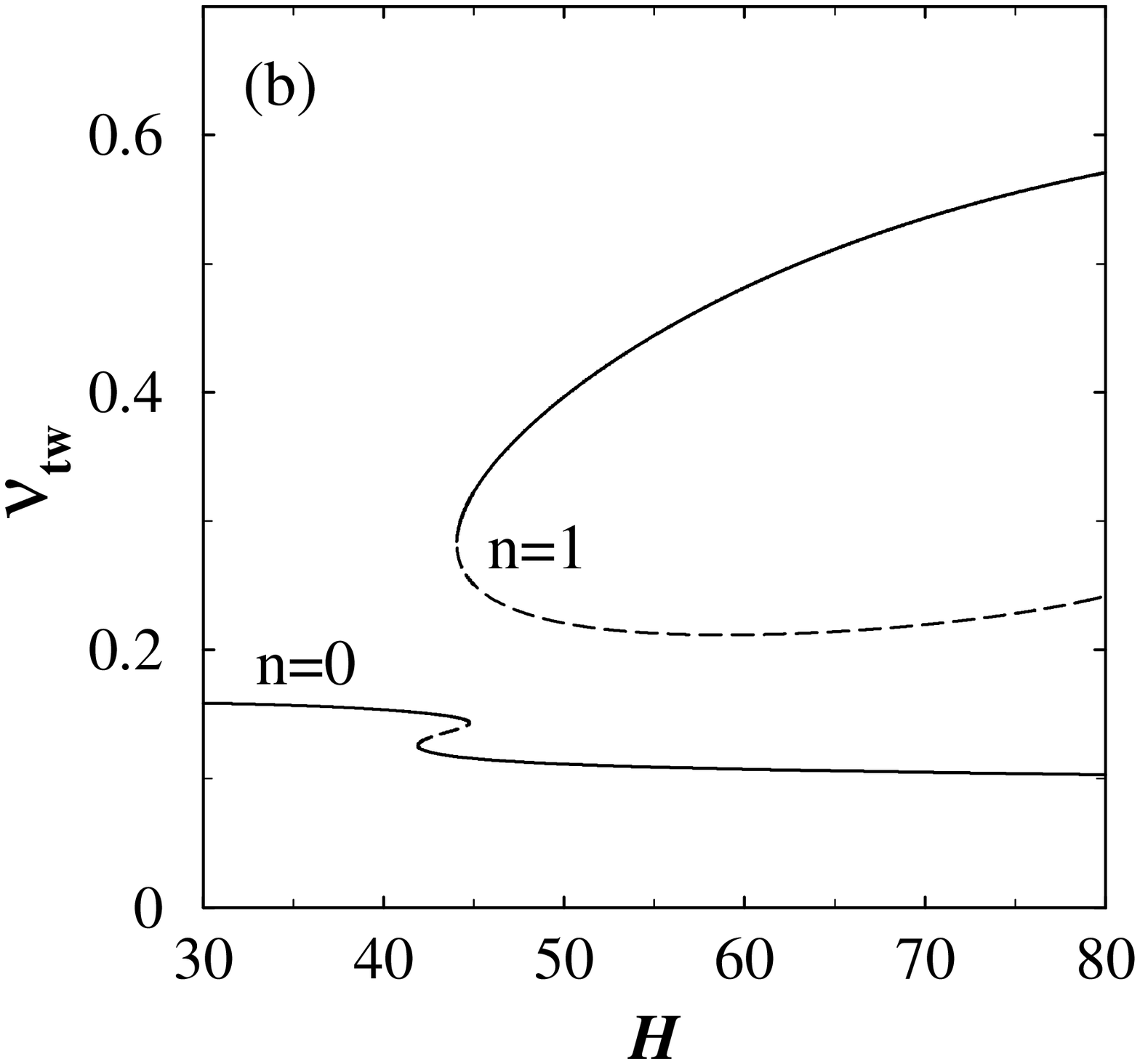}\\
\includegraphics[width=.35\textwidth]{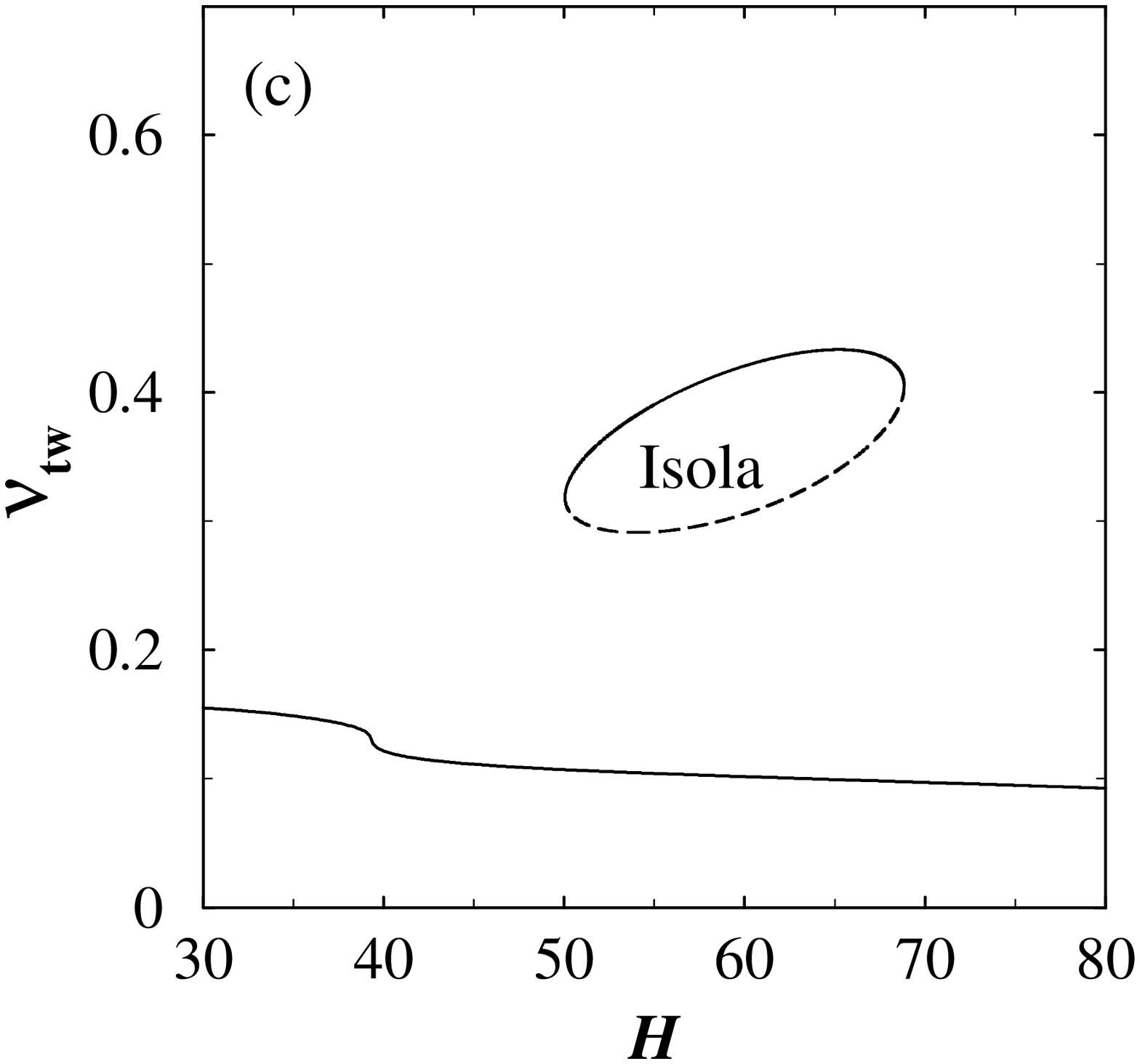}
\includegraphics[width=.35\textwidth]{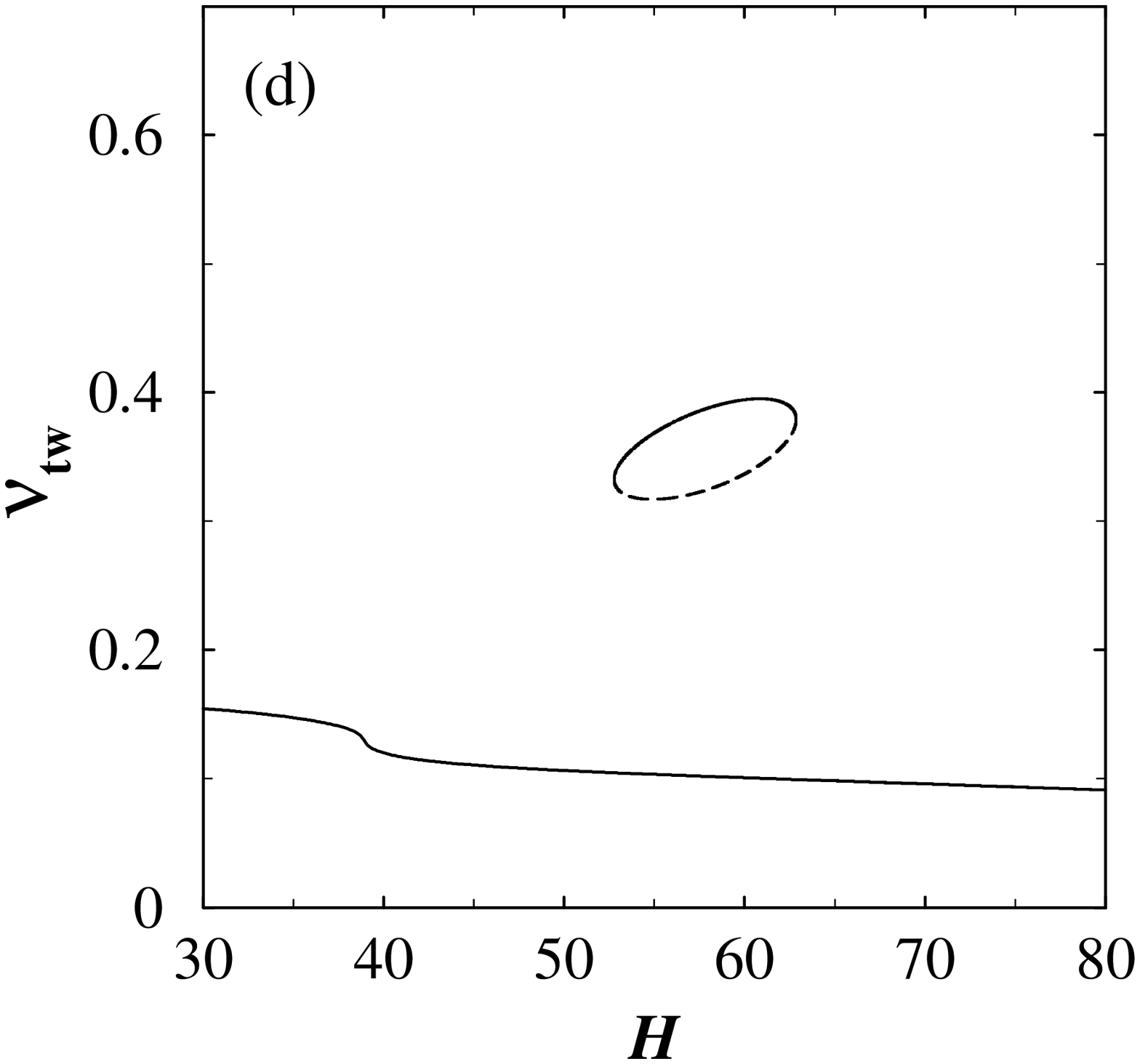}
\end{center}
\caption[]
{
Effect of gravity on the subcritical bifurcations
at $\nu_{av}=0.16$ and $e=0.8$:
(${\bf a}$) $Fr^{-1} = 0$;
(${\bf b}$) $Fr^{-1} =  2\times 10^{-4}$;
(${\bf c}$) $Fr^{-1} = 4\times 10^{-4}$;
(${\bf d}$) $Fr^{-1} = 4.25\times 10^{-4}$.
The {\it stable} ({\it unstable})
bifurcation branch is denoted by solid (dashed) line
}
\label{fig:fig6}
\end{figure}

We have seen that the shear-banding instabilities in granular Couette flow also
lead to subcritical pitchfork bifurcations
that occur at lower mean densities (cf. Fig. \ref{fig:fig2}$a$).
The effect of gravity on such subcritical bifurcations
is shown in Fig. \ref{fig:fig6}, with parameter values 
set to $\nu_{av}=0.16$ and $e=0.8$.
Note that the smooth branch in Fig. \ref{fig:fig6}($b$) 
now contains two {\it saddle-nodes}, joined by 
a `hysteretic' branch; such `hysteretic' jumps,
however, disappear when the gravitational strength is strong
(i.e. large $Fr^{-1}$) as seen in Fig. \ref{fig:fig6}$c$.
From Figs. \ref{fig:fig6}($c$) and \ref{fig:fig6}($d$)
we observe that the size of the isola 
diminishes with increasing $Fr^{-1}$, eventually disappearing
at $Fr^{-1}\sim 4.4\times 10^{-4}$ for this parameter combination.
Hence, under strong gravitational strength, the surviving attractor
in the phase-space is the one that corresponds to a solution
with a plug near the bottom wall and a dilute almost uniformly sheared
layer near the top-wall.

Returning back to the canonical bifurcation diagrams for
universal unfolding in Fig. \ref{fig:fig5},
we observe that the bifurcation diagram for zone $3$
is similar to that in Fig. \ref{fig:fig6}($b$).
Again redefining our order-parameter as $\Phi=\nu_{bw}\equiv \nu(-1/2)$
and then redrawing
the bifurcation diagrams of Fig. \ref{fig:fig6},
we  get back the canonical set for zone $4$.
It is now clear that the inclusion of gravity 
naturally recovers the two `hysteretic'
bifurcation diagrams for zone $3$ and zone $4$
in Fig. \ref{fig:fig5}.
Thus, all possible forms of imperfect bifurcation
scenarios can be realized in the present bifurcation problem.
We can conclude that the effect of gravity on
the granular Couette flow truely belongs to the class of
the `universal' unfolding of pitchfork bifurcations.

\subsection{Comparison with Experiments and Discussion}

In the regime of dense flows with low shear rates,
the experiments and molecular dynamics simulations \cite{SB1984,TG1991}
suggest that the shearing mainly occurs in a thin-layer 
near the top wall, with most of the material remaining
almost undeformed near the bottom wall. Recall that
in most shear-cell experiments
the Froude number is of order $10$ (i.e. $Fr^{-1}\sim 0.1$).
This is precisely what we have predicted from our analysis:
at large values of $Fr^{-1}$ there exists
a unique solution with a plug near the bottom wall.
Such large values of $Fr^{-1}$ in earth-bound shear-cell experiments
clearly preclude the emergence of other solutions that have plugs
near the top wall or somewhere in between.
Under {\it micro-gravity} conditions (i.e. at small  $Fr^{-1}$), however,
it would be possible to observe
such  multiple-plugs, `floating' within  the Couette gap.
Thus, we can  conclude that
the gravitational Couette flow admits only one solution,
having a plug near the bottom wall and a shear-band near the top wall 
\cite{Alam1998a,Alam2004}.
We have established that the shearband formation in
gravitational plane Couette flow is tied to the
shearbanding instabilities in `uniform shear flow'
via the route of the universal unfolding of pitchfork bifurcations.

\begin{flushleft}
{\large\bf Acknowledgements}
\end{flushleft}
{\it Financial support and computational facilities from  the Jawaharlal Nehru Center
for Advanced Scientific Research are acknowledged.
Partial financial support from the AvH Foundation (to attend the
STAMM'04 meeting) is also acknowledged.}

\end{document}